\PassOptionsToPackage{x11names}{xcolor}
\documentclass[twocolumn, tighten]{aastex631}
\usepackage{starType}
\newcommand*{\SAX}{\source{SAX}{J1808.4}{-3658}}

\newcommand*{\MdotEdd}{\ensuremath{\dot{M}_{\mathrm{Edd}}}}
\newcommand*{\Zcno}{\ensuremath{Z_{\mathrm{CNO}}}}
\newcommand*{\meanAcno}{\ensuremath{\bar{A}_{\mathrm{CNO}}}}
\newcommand*{\thalf}[1][]{\ensuremath{t_{1\!/\!2,\mathrm{#1}}}}
\newcommand*{\ndens}[1][]{\ensuremath{n_{\mathrm{#1}}}}
\newcommand*{\yign}{\ensuremath{y_{\mathrm{ign}}}}
\newcommand*{\tign}{\ensuremath{t_{\mathrm{ign}}}}
\newcommand*{\Ltail}{\ensuremath{L_{\mathrm{tail}}}}

\newcommand*{\Etail}{\ensuremath{E_{\mathrm{tail}}}}
\newcommand*{\ttail}{\ensuremath{t_{\mathrm{tail}}}}
\newcommand*{\Lpeak}{\ensuremath{L_{\mathrm{peak}}}}

\newcommand*{\Lcno}{\ensuremath{L_{\mathrm{CNO}}}}
\newcommand*{\Lnuc}{\ensuremath{L_{\mathrm{nuc}}}}
\newcommand*{\Lacc}{\ensuremath{L_{\mathrm{acc}}}}
\newcommand*{\timescale}[1]{\ensuremath{\tau_{\mathrm{#1}}}}
\newcommand*{\tnuc}{\timescale{nuc}}
\newcommand*{\tth}{\timescale{th}}
\newcommand*{\cP}{\ensuremath{c_{\!P}}}

\newcommand{\figcolwidth}{244bp}

\begin{document}

\title{Weak X-ray Bursts on Slowly Accreting Neutron Stars}

\correspondingauthor{Sierra Casten}
\email{castensi@msu.edu}

\author[0000-0003-3872-1703]{Sierra Casten}
\affiliation{Michigan State University, East Lansing, MI 48824, USA}

\author[0000-0002-5211-2122]{Simon Guichandut}
\affiliation{Department of Physics and Trottier Space Institute, McGill University, Montreal, QC H3A 2T8 Canada}

\author[0000-0002-6335-0169]{Andrew Cumming}
\affiliation{Department of Physics and Trottier Space Institute, McGill University, Montreal, QC H3A 2T8 Canada}

\author[0000-0003-3806-5339]{Edward F. Brown}
\affiliation{Michigan State University, East Lansing, MI 48824, USA}
\affiliation{Facility for Rare Isotope Beams, Michigan State University, East Lansing, MI 48824 USA}

\begin{abstract}

Nearly all observed thermonuclear X-ray bursts are thought to be triggered by the thermally unstable triple-alpha process. Unlike in accreting white dwarfs, the $\gtrsim10^{8}\vsk\Kelvin$ envelope temperature forces hydrogen burning to proceed via the $\beta$-limited, thermally stable hot CNO cycle. Recent observations of weak X-ray bursts from \SAX\ occurring within 1--3 days of the onset of an accretion outburst have raised the question that these bursts were triggered by thermally unstable hydrogen ignition, analogously to classical novae. Using the stellar evolution code \mesa, we explore the unstable ignition of hydrogen on slowly accreting neutron stars. For solar metallicities, the burning is insufficiently vigorous to launch convection and the burst rise is on the envelope thermal timescale, with the hydrogen being consumed over hours. For elevated metallicities ($Z\gtrsim0.06$), however, the initial proton captures onto CNO nuclei heat the envelope enough to drive convection and produce a sharp peak in the luminosity. This peak can match that of the \SAX\ burst, for sufficient enrichment. Following this peak, the hydrogen burning stabilizes at a rate set solely by the ignition depth and CNO abundance in the accreted matter. This quasi-steady-state burning produces an extended tail of elevated emission that lasts until the hydrogen is exhausted. Observations of the tail's luminosity and duration measure the hydrogen and metallicity abundance of the accreted material, while the distance-independent ratio of peak-to-tail luminosities would suggest the presence of significant metallicity gradients prior to convection.

\end{abstract}

\section{Introduction}\label{s.introduction}

Type I X-ray bursts are thermonuclear runaways that occur on the surface of accreting neutron stars. These bursts are characterized by a sudden increase in flux and typically last from tens to hundreds of seconds \citep[for a review, see][]{Strohmayer2006New-views-of-th,Galloway2021Thermonuclear-X}. The cause of the bursts is a thin-shell instability \citep{hansen75:_thin}, in which fuel accumulates until an upwards fluctuation in temperature causes the local heating rate from nuclear reactions to increase faster than the local cooling rate from thermal conduction. This instability drives a thermonuclear runaway that consumes the available fuel, and the cycle then repeats.

One of the main parameters that determine the behavior of the burst is the accretion rate $\dot{M}$ \citep{taam79:_therm, fujimoto81:_shell_x, Cumming2004Theory-vs-obser}. 
The behavior has been mapped out using a mixture of semi-analytical calculations, in which the heating and cooling timescales were compared to determine stability, and single- and multi-zone models of the envelope. For a thorough review of these estimates, see \citet{Galloway2021Thermonuclear-X}. Roughly speaking, for $\dot{M}\gtrsim 0.01\,\MdotEdd$, the $\beta$-limited hot CNO cycle stably burns H to He \citep{Hoyle1965Report-on-the-P}, where $\MdotEdd\approx \val{\sci{2}{-8}}{\Msunperyr}$ denotes the rate at which the accretion luminosity reaches the Eddington limit.
In this cycle, the rate at which H is consumed is set by the temperature-insensitive $\beta$-decay lifetimes of \oxygen[14] and \oxygen[15], and so the timescale over which H is depleted is solely set by the ratio of H to CNO. At $\dot{M}\gtrsim0.1\,\MdotEdd$, He ignites via the triple-$\alpha$ process before H is exhausted. This results in a mixed H/He burst, in which the protons are captured onto seed nuclei formed from $\alpha$-captures \citep{schatz.aprahamian.ea:endpoint}. The profile shape of the light curve, characterized by a slower rise time ($\approx 5$–$10$ s), is dictated by the H/He ratio \citep{Fisker2008Explosive-Hydro}, while the long decaying tail ($\approx$ minutes) is sensitive to the efficiency of nuclear burning at various waiting points in the rp-process, as explored by \citet{woosley.heger.ea:models} using multi-zone models. Additionally, the metallicity of the accreted material plays a key role in shaping burst properties, influencing recurrence times, peak luminosities, and nucleosynthesis pathways \citep{Jose2010}. When $\dot{M}\lesssim\ 0.1 \,\MdotEdd$ and the surface temperature is $\gtrsim 8\times10^{7} \mathrm{K}$, there is sufficient time for the hot-CNO cycle to completely consume the available H, so that the triple-$\alpha$ process ignites in a nearly pure He envelope. In contrast to the mixed H/He bursts, pure-He bursts are more intense, often Eddington-limited, and are shorter in duration.

At $\dot{M}\lesssim\ 0.01 \,\MdotEdd$, the envelope temperature may be cool enough, $\lesssim \val{\sci{8}{7}}{\K}$, that the CNO cycle is no longer $\beta$-limited and therefore becomes thermally unstable. If the unstable H burning heats the envelope sufficiently, He will also ignite and produce a mixed H/He burst. At even lower accretion rates, the heating from unstable H burning may not be sufficient to ignite He. Under these conditions, a weak burst powered solely by H burning would result. Although there have been several theoretical studies of burning at low $\dot{M}$ \citep[see, e.g.,][]{Cumming2004Theory-vs-obser, Peng2007Sedimentation-a, Cooper&Narayan2007Hydrogen-trigge}, observational examples of H-triggered bursts have been elusive. \citet{Casten2023Hydrogen-trigge} observed such a candidate burst in 2019 from the accreting millisecond pulsar \SAX\ with the \emph{Neutron Star Interior Composition Explorer} (\nicer). An archival search by \citet{Casten2023Hydrogen-trigge} found a similar burst from \SAX\ that was observed by the \emph{Rossi X-ray Timing Explorer} (\rxte) in 2005.

\citet{Casten2023Hydrogen-trigge} argued that these weak bursts were consistent with being solely due to unstable H burning. They noted that the cool envelope temperature required for unstable H ignition was consistent with both observed weak bursts occurring 1--3 days after the onset of active accretion, and also that the existence of a long, steady tail of extended emission following the peak was consistent with stable CNO burning \added{\citep[see][]{Peng2007Sedimentation-a}}. Adjusting for an updated distance of $\val{3.3}{\kpc}$ \citep{Goodwin2019A-Bayesian-appr}, \citeauthor{Casten2023Hydrogen-trigge} found a peak luminosity of $\val{9.1\times10^{36}}{\ergspersecond}$ and a tail luminosity of $\approx\val{7.9\times10^{34}}{\ergspersecond}$. 

At low accretion rates, $\dot{M}\lesssim\ \val{10^{-2}}{\MdotEdd}$, sedimentation of heavy elements in the envelope \citep{bildsten92,wallace.woosley.ea:thermonuclear, Peng2007Sedimentation-a} could potentially enhance the CNO abundance at the base of the accreting envelope and lead to a more energetic burst. The role of enhanced CNO abundances in driving vigorous hydrogen burning is well established in models of classical novae (\citealt{Starfield1993Theory-and-Obse}; see also \citealt{Denissenkov2013MESA-Models-of-} and references therein), though the exact mechanisms responsible for this enhancement is not well understood. Mixing processes, such as convective overshooting and Kelvin-Helmholtz instabilities, are thought to play a key role in enriching the accreted envelope with heavier elements from the underlying white dwarf \citep{Casanova2018Two-dimensional, Casanova2011Kelvin-Helmholt}.

\citet{Peng2007Sedimentation-a} studied this effect in a simplified one-zone model, but it has not yet been included in multi-zone simulations. In this preliminary study, we do not include sedimentation directly, but instead study the effect of enhanced metallicity on the properties of the burst by considering a range of accreted metallicities. We use the open-source stellar evolution code \mesa\ (\citealt{Paxton2011}, \citealt{Jermyn2023Modules-for-Exp} and references therein) to explore the unstable ignition of hydrogen at low mass accretion rates for a range of metallicities in the accreted fuel. As shown by the one-zone calculations of \citet{Peng2007Sedimentation-a} and two-zone model of \citet{Cooper&Narayan2007Hydrogen-trigge}, after the initial peak in luminosity, there is a long tail of enhanced luminosity powered by stable hydrogen burning via the hot CNO cycle. This burning continues until it consumes the accumulated hydrogen, at which point the envelope cools and the luminosity drops.

We revisit quasi-steady-state H burning via the hot CNO cycle in \S~\ref{s.cno-cycle} and provide estimates for the luminosity and duration of the tail. We then describe our \mesa\ models in \S~\ref{s.burst-model}. In this preliminary study we do not include sedimentation but instead model a range of metallicities $Z=0.01\textrm{--}0.30$, which is a larger range than expected, but allows us to fully explore the effect of enhanced CNO. For comparison, \citet{Peng2007Sedimentation-a} estimated that sedimentation in a hydrogen-carbon mixture could increase the abundance of \carbon\ at the base of the accreted layer from 0.02 to 0.23. Because the weak bursts were observed shortly after the onset of accretion, we do not run our simulations through several bursts but instead focus on the first burst following the onset of accretion. We show (\S~\ref{s.unstable-burning}) how the CNO abundance affects the ignition depth, peak luminosity, and burst lightcurve morphology, and we briefly compare our burst lightcurves and fluences with those of the 2019 burst observed from \SAX. We conclude in \S~\ref{s.discussion} with a summary, a brief discussion of uncertainties, and an outlook on future efforts.

\section{Burning via the Hot CNO cycle}\label{s.cno-cycle}

As shown by \citet{Peng2007Sedimentation-a}, following the initial triggering of the burst by the cold CNO cycle reactions, the temperature increases enough that the burning enters the hot CNO cycle and becomes steady with the net rate set solely by the temperature-independent $\beta$-decays of \oxygen[14] and \oxygen[15] with half-lives $\thalf[O14] = \val{70.64}{\second}$ and $\thalf[O15] = \val{122.24}{\second}$, respectively \citep[JINA REACLIB;][]{Cyburt2010}. In this steady-state burning, all CNO nuclides are converted into \oxygen[14] and \oxygen[15], with the number densities $n_{\mathrm{O14}}$ and $n_{\mathrm{O15}}$ being in the ratio $n_{\mathrm{O14}}/n_{\mathrm{O15}} = \thalf[O14]/\thalf[O15]$.   In this regime, hydrogen is depleted at a rate
\begin{equation}
\label{e.H-depletion}
\DDt{\ndens[H]} = -4\ndens[CNO]\frac{\ln 2}{\thalf[O14]+\thalf[O15]}.
\end{equation}
Here $\ndens[CNO]$ denotes the number density of all CNO nuclides; during hot CNO burning, $\ndens[CNO]=n_{\mathrm{O14}}+n_{\mathrm{O15}}$. Thus, if H consumption proceeds via steady-state hot CNO burning in the burst tail, the layer of hydrogen accumulated at the start of the burst will be depleted in a time
\begin{equation}
\label{e.depletion-time}
t_{d} = \frac{\ndens[H]}{\ndens[CNO]} \frac{\thalf[O14]+\thalf[O15]}{4\ln2},
\end{equation}
which depends solely on the ratio of hydrogen to CNO nuclides at the onset of burning.

The mean mass of the CNO nuclides in the accreted fuel is $\meanAcno = 14.56$ and the mass fraction of CNO elements for a given overall metallicity $Z$ is $\Zcno = 0.6677\,Z$ \citep{Asplund2005The-Solar-Chemi}. As a consequence, the ratio of hydrogen (with mass fraction $X$) to CNO nuclides is
\begin{equation}
\label{e.HCNO}
\frac{\ndens[H]}{\ndens[CNO]} = \frac{\meanAcno\,X}{\Zcno} = 21.80\frac{X}{Z}.
\end{equation}
Substituting Eq.~(\ref{e.HCNO}) into Eq.~(\ref{e.depletion-time}) along with \thalf[O14] and \thalf[O15] yields
\begin{equation}
\label{e.num-depletion-time}
t_{d} = \val{\sci{1.1}{5}}{\second} \left(\frac{X}{0.7}\right)\left(\frac{0.01}{Z}\right).
\end{equation}
\citet{Lampe2016THE-INFLUENCE-O} derived a similar expression; their prefactor differs from ours because they took the metallicity to entirely reside in CNO isotopes, i.e., $Z=\Zcno$.

The production of each \helium\ nucleus is accompanied by a release of heat, $Q=\val{(26.73-3.22)}{\MeV} = \val{24.7}{\MeV}$, in which the first term is the mass difference and the second term accounts for the neutrino losses\footnote{We adopt here the mean neutrino energy used by \mesa, $\bar{E}_{\nu}=\val{3.22}{\MeV}$, which is larger than the value of $\val{2.03}{\MeV}$ used by \citet{wallace81:_explos}.}.
The luminosity from CNO burning in the burst tail is therefore
\begin{eqnarray}
\Lcno &=& \frac{4\pi R^{2}\yign X Q}{4\mb t_{d}} \label{e.LCNO} \\
 &=& \val{\sci{4.8}{34}}{\ergspersecond} \left(\frac{\yign}{\val{\sci{3}{7}}{\columnunit}}\right)\left(\frac{Z}{0.02}\right).\nonumber
\end{eqnarray}
Here $\mb$ is the atomic mass unit, and $\yign$ is the column depth of the base of the accreted H-rich layer, with column depth defined as $y = \int\rho\,\dif r \approx \Delta M_{\mathrm{accr}}/(4\pi R^{2})$. For $R=\val{12}{\km}$ the ignition mass is $\Delta M_{\mathrm{ign}} = \val{\sci{2.7}{-13}}{\Msun}\left(\yign/\val{\sci{3}{7}}{\columnunit}\right)$ with a recurrence time $t_{r} = \val{24.0}{\hour}\left(\yign/\val{\sci{3}{7}}{\columnunit}\right)$ for $\Mdot=\val{10^{-10}}{\Msunperyr}$.
Note that \Lcno\ depends solely on \yign\ and the metallicity $Z$ of the accreted layer; in particular, it does not depend on the mass fraction of hydrogen $X$.

For comparison, the luminosity produced by burning hydrogen at the rate at which it accretes is
\begin{eqnarray}
L_{\mathrm{SS}} &=& \frac{\dot{M}XQ}{4\mb}\nonumber\\
 &=& \val{\sci{2.5}{34}}{\ergspersecond}\left(\frac{X}{0.7}\right)\left(\frac{\dot{M}}{\val{10^{-10}}{\Msunperyr}}\right).
\label{e.Lsteady}
\end{eqnarray}
For $L_{\mathrm{CNO}} > L_{\mathrm{SS}}$, fuel cannot be replenished quickly enough to sustain $L_{\mathrm{CNO}}$, and so, after a time $t_{d}$, the luminosity drops and the envelope cools on a thermal timescale. If the temperature is too cool, the proton captures in the CNO cycle cease and the envelope settles into a limit cycle.

\section{Numerical models}\label{s.numerical-models}
\label{s.burst-model}

Having described the basic physics of steady-state hot CNO burning following (unstable) H ignition, we now 
model the accretion and ignition. We do this using the open-source stellar evolution code \mesa, version
r23.05.01  \citep{Paxton2011, Paxton2013, Paxton2015, Paxton2018, Paxton2019, Jermyn2023}, and compiled with the \MESA~\code{SDK}, version \code{x86-linux-23.7.3} \citep{Townsend2024MESA-SDK-for-Li}. The microphysics employed by \MESA\ are as follows. The equation of state (EOS) is a blend of the OPAL \citep{Rogers2002}, SCVH \citep{Saumon1995}, FreeEOS \citep{Irwin2004}, HELM \citep{Timmes2000}, PC \citep{Potekhin2010}, and Skye \citep{Jermyn2021} EOSs. Radiative opacities are primarily from OPAL \citep{Iglesias1993,Iglesias1996}, with low-temperature data from \citet{Ferguson2005} and with the high-temperature, Compton-scattering dominated regime treated following \citet{Poutanen2017}. Electron conduction opacities are from \citet{Cassisi2007} and \citet{Blouin2020}. Nuclear reaction rates are taken from JINA REACLIB \citep{Cyburt2010} and NACRE \citep{Angulo1999} and supplemented by additional tabulated weak reaction rates \citep{Fuller1985, Oda1994, Langanke2000}. Screening is included via the prescription of \citet{Chugunov2007}. Thermal neutrino loss rates are included using the prescription of \citet{Itoh1996}.

We construct the envelope model using the \MESA\ test suite case \code{make\_env} with a neutron star mass $M=1.4\,\Msun$, radius\footnote{We adopt here  $R=\val{12}{\km}$, which is consistent with recent \nicer\ determinations of the radius of the high-mass pulsar \source{PSR}{J0740}{+6620} \citep{Dittmann2024A-More-Precise-,Salmi2024The-Radius-of-t}. For this initial, exploratory, study, we do not sample over the distribution of neutron star radii.} $R=12\,\km$, and luminosity $L_{b}=\val{10^{33}}{\ergspersecond}$. We set the initial composition of the envelope to \iron\ with a mass $\val{\sci{1.8}{24}}{\gram}$, so that the base of the computational domain is initially at a column of $10^{11}\,\columnunit$. We then accrete H-rich matter at a rate $\Mdot = \val{10^{-10}}{\Msunperyr}$. The accretion rate is low in order to ensure a sufficiently cool envelope for the CNO cycle to be thermally unstable \citep{Cumming2004Theory-vs-obser, Peng2007Sedimentation-a, Casten2023Hydrogen-trigge}. For this accretion rate, $L_{b}/\Mdot = \val{0.16}{\MeV\,\amu^{-1}}$ ($=\val{\sci{1.5}{17}}{\ergspergram}$), which is consistent with the value needed for unstable H ignition \citep{Peng2007Sedimentation-a}.

The distribution of nuclides in the accreted matter follows the solar system abundances complied by \citet{Asplund2005The-Solar-Chemi}. To follow any potential breakout from the hot CNO cycle into an rp-process \citep{wallace81:_explos}, we use a modified version of the truncated reaction network \code{rp\_153.net}, which is a subset of the full rp-process network, as it only extends to \nickel[56]. This is sufficient for modeling weak bursts, which do not develop a strong rp-process. We checked that the abundances of the heaviest isotopes did not appreciably change during the burst.
To account for species that are not present in the network without renormalizing abundances, we use the default setting (\code{.true.}) for the input flag \code{accretion\_dump\_missing\_metals\_into\_heaviest} and add \cobalt[57], \nickel[57], and \nickel[58] (which is stable) to \code{rp\_153.net}; otherwise there would be spurious heating from the decay of \nickel[56]. In addition, we include \carbon[13] to allow for decays of \nitrogen[13] during the onset of the CNO cycle. Our modified reaction network, \code{rp\_157.net} thus consists of the isotopes shown in Table~\ref{t.rp156}.
\begin{deluxetable}{lr|lr|lr|lr}
\tablewidth{0pt}
\tablecaption{\label{t.rp156}Isotopes in the reaction network \code{rp\_157.net}.}
\tablehead{
\colhead{El.} & \colhead{$A$} & \colhead{El.} & \colhead{$A$} & \colhead{El.} & \colhead{$A$} & \colhead{El.} & \colhead{$A$}
}
\startdata
H  & 1--3     & O  & 13--18 & P  & 26--31 & Ti & 40--47 \\
He & 3,4      & F  & 17--19 & S  & 27--34 & V  & 43--49 \\
Li & 7        & Ne & 18--21 & Cl & 30--35 & Cr & 44--52 \\
Be & 7,8      & Na & 20--23 & Ar & 31--38 & Mn & 47--53 \\
B  & 8,11     & Mg & 21--25 & K  & 35--39 & Fe & 48--56 \\
C  & 9,11--13 & Al & 22--27 & Ca & 36--44 & Co & 51--57 \\
N  & 12--15   & Si & 24--30 & Sc & 39--45 & Ni & 52--58 \\ 
\enddata
\tablecomments{The reaction network also includes free neutrons.}
\end{deluxetable}

In order to better resolve the thermal instability, we reduce the time resolution control \code{time\_delta\_coeff} from 1.0 (the default) to 0.1. We model convection following the mixing length theory of \citet{cox68:_princ_stell_struc} with mixing length parameter $\alpha = 1.5$, and we use the (default) Schwarzschild criterion, rather than Ledoux, to determine convective stability. We find similar results using each criterion but using Schwarzschild improves numerical stability over all metallicities.

\section{Unstable hydrogen burning}\label{s.unstable-burning}

We model accreting neutron star envelopes with metallicities $Z$ ranging from 0.01 to 0.30. In all cases, we set the hydrogen mass fraction $X = 0.70$ and the helium mass fraction $1-X-Z$. We defer the exploration of more physically realistic conditions, such as would occur under sedimentation, to a subsequent paper. We first explore the onset of unstable hydrogen burning for two representative cases, $Z=0.04$ and $Z=0.10$. For $Z=0.04$, the envelope does not become convective during the burst, and as a result the burst rise is lengthy; in contrast, for $ Z=0.10$ the burst rise is more violent and leads to the growth of a convective zone that rapidly transports heat to the surface. In both cases, following this peak the bursts exhibit a long tail of quasi-steady hydrogen burning via the hot CNO cycle. This continues until the hydrogen is depleted, at which point the envelope cools. After exploring these two cases in detail, we then examine features of the hydrogen burning over the range of sampled metallicities.

\subsection{The onset of unstable hydrogen burning}
\label{s.onset-hydrogen-burning}

\begin{figure}[htbp]
\centering
\includegraphics[width=\figcolwidth]{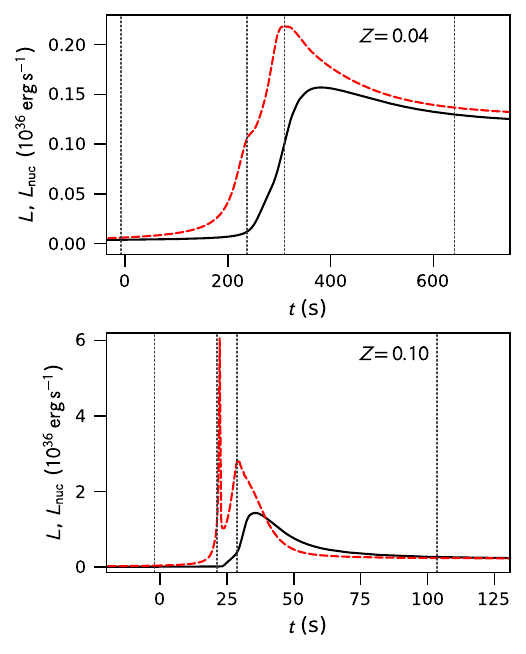}
\caption{\label{f.lightcurves} Surface luminosity (\emph{solid black}) and nuclear luminosity (\emph{dashed red}) during the onset of unstable hydrogen burning and the evolution of the burning to a tail of quasi-steady burning. Models with $Z = 0.04$ (\emph{top}) and 0.10 (\emph{bottom}) are shown. The vertical dotted lines indicate where we sample the isotopic distribution in the accreted and burning neutron star envelope.
}
\end{figure}

Figure~\ref{f.lightcurves} illustrates the onset of the instability and evolution of the burst for $Z = 0.04$ and $Z = 0.10$. We show both the surface luminosity $L$ (\emph{solid black curves}) and the integrated nuclear heating $\Lnuc = \int \epsnuc\,\dif m$ (\emph{dashed red curves}), which is stored in \mesa\ as \code{log\_power\_nuc\_burn}\footnote{In a quirk of nomenclature, the parameter \code{log\_Lnuc} subtracts off the heating from photodissociation reactions and so does not represent the full nuclear luminosity.}. The thin vertical dashed lines denote times for which we shall show isotopes in the accreted envelope. A feature in the nuclear luminosity is the presence of an inflection, which steepens into a spike, followed by a rise to maximum. Similar behavior in the nuclear heating is observed in models of classical novae \citep{Prialnik1979The-evolution-o}.

\begin{figure*}[tbp]
\includegraphics[width=\textwidth]{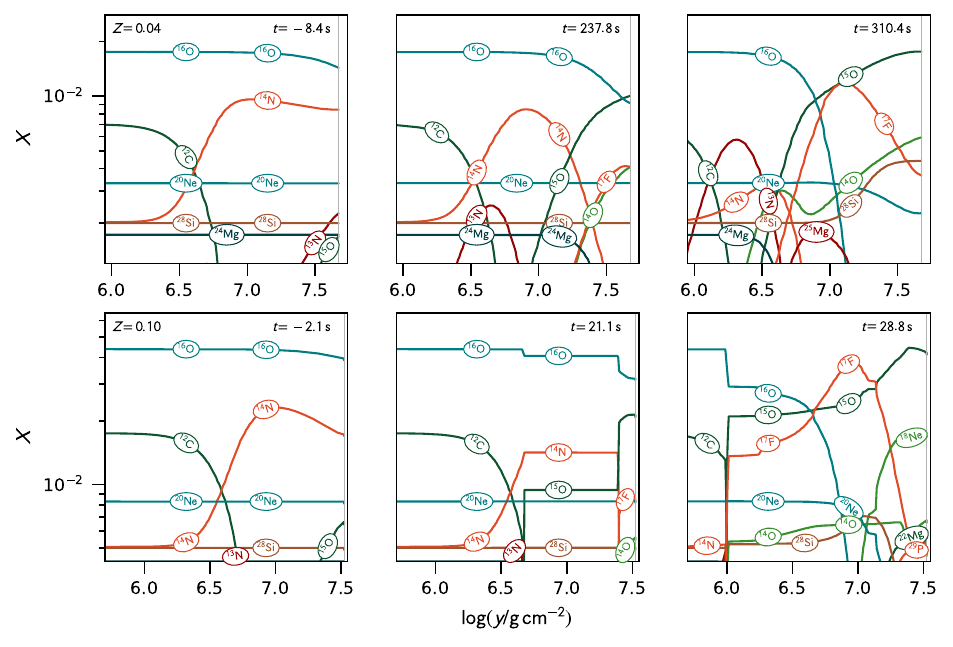}
\caption{\label{f.snapshots}
Mass fractions as functions of column depth $y$ for accretion with $Z=0.04$ (\emph{top row}) and $Z=0.10$ (\emph{bottom row}). The columns, from left to right, are a time-ordered are correspond to the first three vertical lines depicted in Fig.~\protect\ref{f.lightcurves}. The lines terminate right at the base of the accreted layer. The isotope \iron\ is not shown to avoid clutter.
}
\end{figure*}

Figure~\ref{f.snapshots} displays the evolution of the composition for these cases. The top row shows mass fractions of isotopes as functions of column depth in the accreted envelope for $Z = 0.04$; the bottom, for $Z=0.10$. Going from left to right, the plots represent snapshots corresponding to the first three thin vertical dotted lines in Fig.~\ref{f.lightcurves}. The first snapshot (Fig.~\ref{f.snapshots}) occurs when the local nuclear heating timescale, $\tnuc = \cP T/\epsnuc$, is comparable to the local thermal time \citep{henyey69},
\begin{equation}\label{e.thermal-timescale}
 \tth = \frac{1}{4}\left[\int_{0}^{\yign} \left(\frac{3\cP\kappa}{4acT^{3}}\right)^{1/2} \,\dif y\right]^{2},
\end{equation}
at the base of the accreted layer. Here $\cP$ is the specific heat at constant pressure and $\kappa$ is the opacity.

Before the onset of the burst, the reactions $\carbon(\pt,\gamma)\nitrogen[13](e^{+}\nu_{e})\carbon[13](\pt,\gamma)\nitrogen[14]$ have stably depleted \carbon, and built up the abundance of \nitrogen. The onset of the burst is heralded by the rapidly increasing $\nitrogen(\pt,\gamma)\oxygen[15]$ as the envelope heats. The nuclear timescale $\tnuc$ does not, however, decrease faster than $\tth$. As a consequence, the luminosity rises on a thermal timescale: the envelope does not become convective. Once \nitrogen\ has been converted to \oxygen[15] (Fig.~\ref{f.snapshots}, \emph{top middle}), the relatively long half-life of \oxygen[15] slows the heating rate and creates the inflection in \Lnuc\ (Fig.~\ref{f.lightcurves}, \emph{top}).

At this point, $\oxygen[16](\pt,\gamma)\fluorine[17]$ reactions open additional channels of the hot CNO cycle. At cooler temperatures, \fluorine[17] ($\thalf[F17] = \val{64.49}{\second}$) decays and leads to $\oxygen[17](\pt,\alpha)\nitrogen$, while at higher temperatures we instead have $\fluorine[17](\pt,\gamma)\neon[18](e^{+}\nu_{e})\fluorine[18](\pt,\alpha)\oxygen[15]$. Concurrently, $\oxygen[15](e^{+}\nu_{e})\nitrogen[15](\pt,\alpha)\carbon(\pt,\gamma)\nitrogen[13]$ complete the CNO cycle; the higher temperatures now favor, however, $\nitrogen[13](\pt,\gamma)\oxygen[14]$ instead of $\nitrogen[13](e^{+}\nu_{e})\carbon[13]$. The burning then proceeds to a maximum in \Lnuc\ (Fig.~\ref{f.snapshots}, \emph{top right}) at $t=\val{310.4}{\second}$, at which time \oxygen\ has been consumed and there is a buildup of \oxygen[14] and \oxygen[15]. There are some light element captures on heavier nuclei, as noted by the evolution of \magnesium[25] and the buildup of \silicon[28]. The peak luminosity remains well below the accretion luminosity $\Lacc \approx \val{10^{36}}{\ergspersecond}$.

\begin{figure*}[tp]
\centering
\begin{interactive}{animation}{Z0.10_composition.mov}
\href{https://sites.google.com/view/sierramcasten/current-and-past-projects/current-mesa-x-ray-bursts?authuser=0}{\includegraphics[height=0.5\textheight]{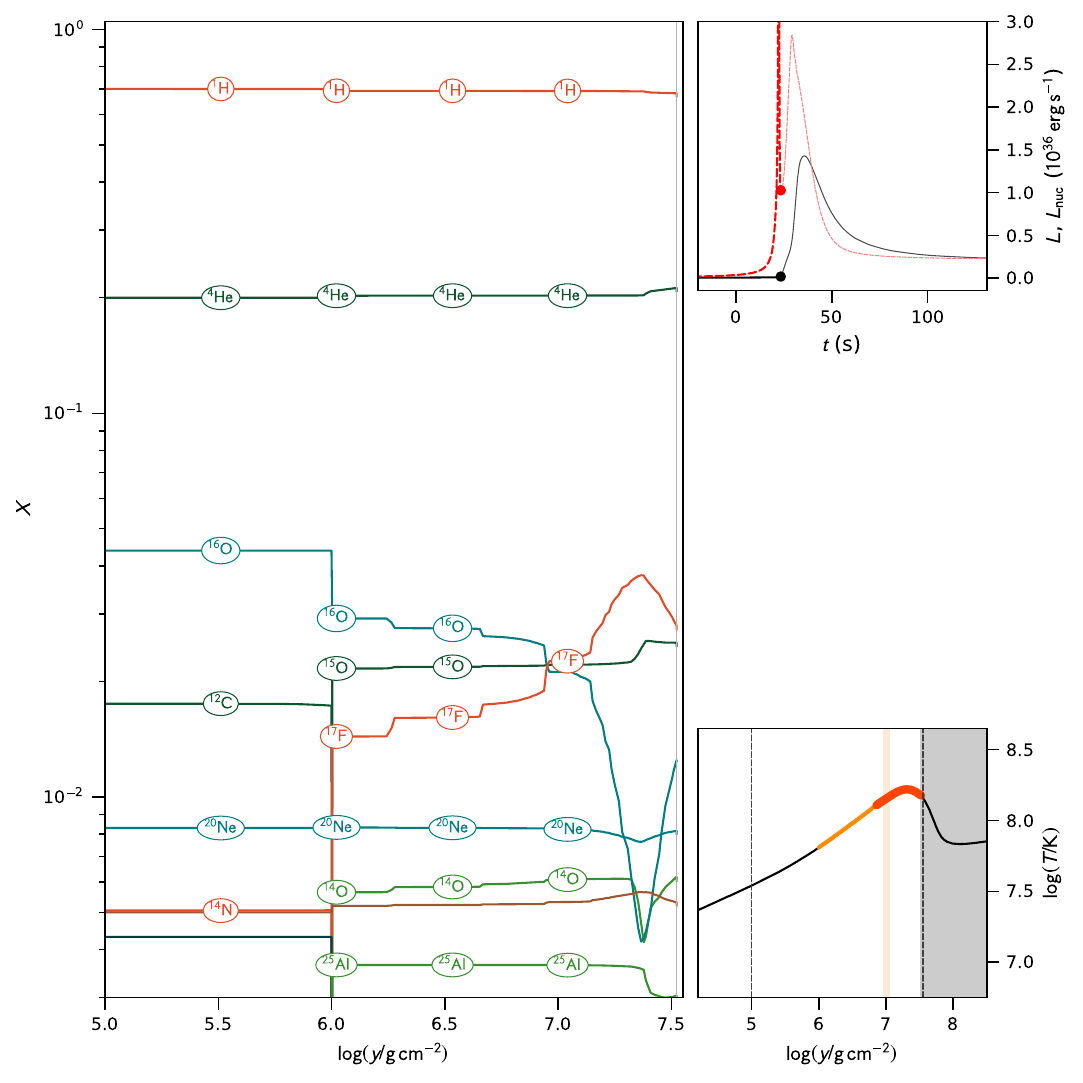}}
\end{interactive}
\caption{\label{f.composition-movie}Evolution of the burst for $Z=0.10$. The \href{https://sites.google.com/view/sierramcasten/current-and-past-projects/current-mesa-x-ray-bursts?authuser=0}{animated figure}, of which the inline figure is a frame, lasts \val{21}{\second} comprises three panels. The main panel at left shows the mass fractions (omitting \iron) in the accreted layer as a function of column; as the animation proceeds the vertical dotted line, marking the base of the accreted layer, sweeps from left to right. The top right panel shows the surface luminosity (\emph{solid black curve}) and integrated nuclear heating (\emph{dashed red curve}); as the animation progresses, dots marking the current time will trace out the luminosity with time (note that the nuclear heating extends beyond plot limit, cf.\ Fig.~\protect\ref{f.lightcurves}, bottom). The bottom right panel shows the temperature as a function of column depth. The vertical dotted lines on the bottom panel indicate the range of column depth displayed in the main panel. The grey shaded region at right is the \iron\ substrate of the initial, relaxed, NS envelope prior to accretion. As the animation progresses, the gray region will displace rightwards. Light tan regions will appear in this panel wherever convection occurs. The temperature curve will turn orange-red where $\epsnuc > 0.05\epsilon_{\mathrm{CNO}}$ and red where $\epsnuc > \epsilon_{\mathrm{CNO}}$.}
\end{figure*}

For $Z=0.10$, the consumption of \carbon\ and buildup of \nitrogen\ (Fig.~\ref{f.snapshots}, \emph{bottom left}) is also thermally stable; in this case, however, the ignition of \nitrogen\ is sufficiently energetic that $\tnuc$ decreases faster than $\tth$ at the base of the envelope. The spike in heating at $t=\val{21.1}{\second}$ (Fig.~\ref{f.lightcurves}, \emph{bottom}) rapidly heats the base of the accreted envelope and drives convection that mixes the deeper part of the envelope and transports heat outwards, which increases $L$. The buildup of \oxygen[15] with its relatively long half-life leads to a rapid, but temporary, fall in heating until $\oxygen(\pt,\gamma)\fluorine[17]$ becomes established. With the fall in heating, convection turns off and does not resume. During the rise to the second peak (Fig.~\ref{f.snapshots} at $t=\val{28.8}{\second}$, \emph{bottom right}) proton captures drive a flow towards \sulfur\ through intermediate nuclides \magnesium[23,25], \aluminum[25], and \phosphorus[29,30]. This burst is sufficiently vigorous that the peak luminosity \Lpeak\ exceeds \Lacc.
A complete evolution of the isotopes, envelope thermal evolution, and lightcurve is contained in the animation linked to Figure~\ref{f.composition-movie}.

\begin{figure}[htbp]
\centering
\includegraphics[width=\figcolwidth]{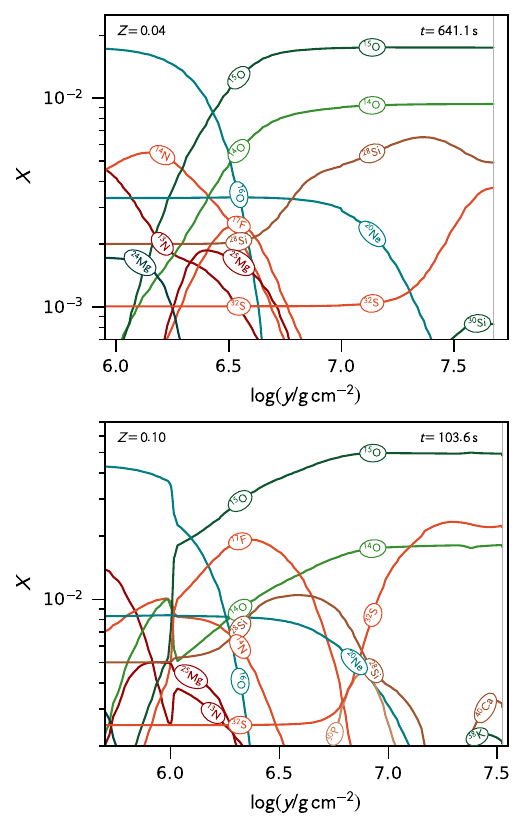}
\caption{\label{f.peak2tail}
Composition at the transition from burst peak to burst tail (the rightmost vertical dotted line in Fig.~\ref{f.lightcurves}), in which H burns via the hot CNO cycle in quasi-steady-state, for $Z = 0.04$ (\emph{top}) and $Z = 0.10$ (\emph{bottom}). The isotope \iron\ is not shown.
}
\end{figure}

For both $Z=0.04$ and $Z=0.10$, the hot CNO cycle becomes established following the peak, and the envelope makes a transition from peak to tail, which we locate by finding where \Lnuc\ is within 5\% of \Lcno\ (Eq.~\ref{e.LCNO}) for \val{20}{\second}. At this point (Fig.~\ref{f.peak2tail}) the mass fractions of \oxygen[14,15] have settled into the ratio set by their half-lives.

\subsection{Burst properties for different metallicities}
\label{s.burst-properties} 

After exploring the onset of unstable hydrogen burning for $Z = 0.04$, which produces a slowing rising luminosity, and for $Z = 0.10$, which launches convection and a strong peak in luminosity, we now present burst models over an ensemble of metallicities. 

\begin{figure}[htbp]
\centering
\includegraphics[width=\figcolwidth]{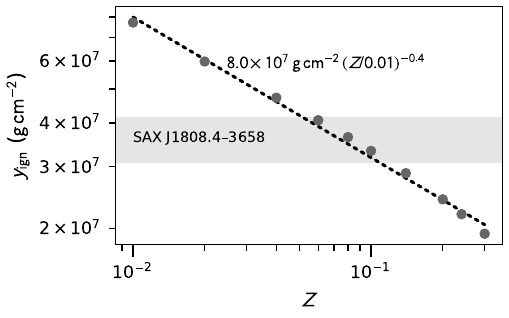}
\caption{\label{fig:column} Ignition column depths (\emph{dots}) as a function of metallicity in the accreted material. These decrease roughly as $\yign\propto Z^{-0.4}$ (\emph{dotted line}). For comparison, we also show the inferred burst ignition depth for \SAX\ (\emph{shaded region}) which covers the range for $\val{1.4}{\Msun} < M < \val{2.0}{\Msun}$. This range of inferred ignition depth is based on the observed fluence for the 2019 \nicer\ burst with $d=\val{3.3}{\kpc}$, $R=\val{12}{\km}$, and anisotropy factor $\xi=1$.
}
\end{figure}

For $Z \ge 0.06$, we locate the time of burst onset, \tign, to be when $(\dif\ln\Lnuc/\dif t)^{-1} < \val{50}{\second}$. For $Z = 0.04$, 0.02, and 0.01, we increase this to 100, 500, and \val{600}{\second}, respectively, since there isn't a strong thermal instability. Figure~\ref{fig:column} shows the corresponding ignition column, $\yign = \tign\times\dot{M}/(4\pi R^{2})$, for different $Z$. We can roughly (maximum error 6\% at $Z=0.30$) fit the ignition column as $\yign \approx \val{\sci{8.0}{7}}{\columnunit}\left(Z/0.01\right)^{-0.4}$. Comparing these ignition columns with those required for unstable triple-$\alpha$ ignition \citep{fujimoto81:_shell_x}, we see that the burst depth is sufficiently shallow to avoid helium ignition. To confirm this, we ran a case with $Z = 0.30$ and hydrogen and helium mass fractions scaled to 0.49 and 0.21, respectively; even with this enhanced helium abundance, there was no significant helium burning.

For comparison, Fig.~\ref{fig:column} also shows (\emph{shaded band}) the inferred ignition column from the 2019 weak burst of \SAX\ observed with \nicer. We estimate this ignition column by integrating the persistent flux, from the onset of accretion following quiescence up to the start of the X-ray burst \citep{Casten2023Hydrogen-trigge}. To convert this integrated flux to an accumulated mass, we adopt a distance $d=\val{3.3}{\kpc}$ \citep{Goodwin2019A-Bayesian-appr} and use two values of neutron star mass, \val{1.4}{\Msun} and \val{2.0}{\Msun}; for both cases we assume\added{\footnote{Although some variation of radius with mass will occur, this variation depends on the specific choice of EOS \citep[for a review, see][]{Watts2016Colloquium:-Mea}, so for simplicity we keep $R$ fixed. Note, however, that the posterior credible interval for $R$ from \nicer\ observations \citep{Dittmann2024A-More-Precise-,Salmi2024The-Radius-of-t} spans over $\val{2}{\km}$, which we do not factor in here.}} $R=\val{12}{\km}$\added{, consistent with recent \nicer\ measurements of the radius of the high-mass pulsar \source{PSR}{J0740}{+6620} \citep{Dittmann2024A-More-Precise-,Salmi2024The-Radius-of-t},} and anisotropy factor $\xi=1$. A higher mass lowers the inferred column, as does a lower distance estimate. Our models with $0.05\lesssim Z\lesssim0.1$ overlap the range inferred from the 2019 burst. Any residual hydrogen from the previous outburst would increase the ignition column, so our estimate is a lower limit on the accreted mass available for the 2019 burst. However, if any additional bursts were missed in the \nicer\ observations prior to the weak burst, which seems plausible, then the estimate for the column would be lowered. In fact, the 2005 \rxte\ burst reported by \citet{Casten2023Hydrogen-trigge} ignited a couple days earlier relative to the \nicer\ burst after the onset of accretion and the column was about a factor of 2 less than the column accumulated prior to the \nicer\ burst. This further supports the need for elevated metallicities.

\begin{figure}[htbp]
\centering
\includegraphics[width=\figcolwidth]{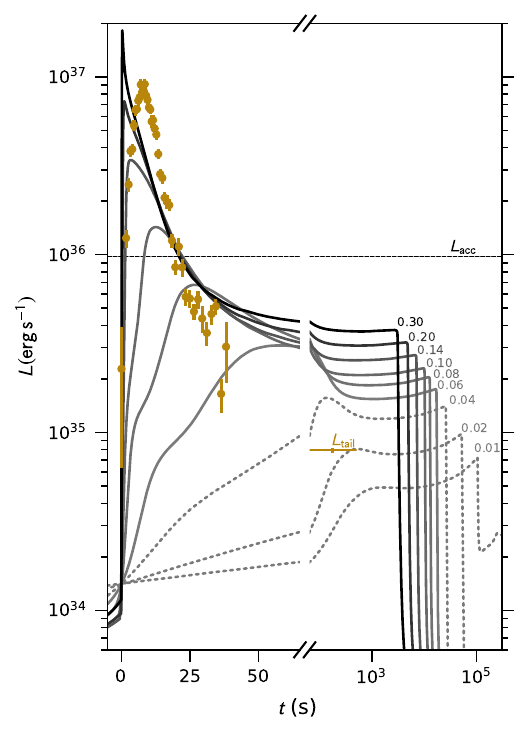}
\caption{\label{f.stacked-bursts} Lightcurves for metallicities from $Z=0.01$ to $Z=0.3$, ordered lightest to darkest, respectively, and numbered by $Z$ at the end of the tail. The lightcurves are aligned so that $L = \val{\sci{1.4}{34}}{\ergspersecond}$ at $t = 0$, and the time scale shifts from linear to logarithmic to follow the transition from peak to tail. Bursts for which no convection develops are indicated with dotted lines and are notable for their long rise. For comparison, we also plot (\emph{gold}) the observations from the 2019 burst, including the measured tail luminosity \citep{Casten2023Hydrogen-trigge}, for $d=\val{3.3}{\kpc}$. Note that the observations have the pre-burst (accretion) emission subtracted. We also indicate (\emph{horizontal dashed line}) the (Newtonian) accretion luminosity.}
\end{figure}

Figure~\ref{f.stacked-bursts} shows the burst lightcurves. The time axis is split, with linear scaling during the initial rise and logarithmic scaling during the long tail. The bursts segregate into two cases, \emph{slow} and \emph{fast}. For the slow bursts with $Z < 0.06$ (\emph{dotted lines}), the burning is not sufficiently vigorous to launch convection, so the rise occurs on the thermal diffusion timescale. For $Z \ge 0.06$, a convective zone briefly develops during the burst rise and produces a more pronounced peak. The contrast between peak and tail increases with $Z$ as the initial spike in heating and resulting convection becomes stronger. At higher $Z$, there is sufficient \nitrogen\ to rapidly heat the envelope and drive convection. The resulting enhanced heat transport produces a sharp, rapid rise in the surface luminosity. Producing a strong peak thus requires a high concentration of CNO elements at the base of the accreted layer. Note, however, that only a small amount of hydrogen is consumed in the peak; for example, the equivalent $\alpha$-value, defined as the ratio of the integrated accretion luminosity prior to the burst to the integrated luminosity of the peak, is $\alpha \gtrsim 10^{3}$ for $Z = 0.10$, far in excess of $\alpha \sim 100$ for typical X-ray bursts.

For comparison, the observed 2019 \nicer\ burst, with accretion luminosity (\emph{horizontal dashed line}) subtracted, is overlaid (\emph{gold points}). This observed burst has \Lpeak\ similar to that for our model $Z=0.20$, albeit with a slower rise. This slower rise could be a consequence of the finite time for the burning front to propagate across the surface \citep[see, e.g.,][]{Cavecchi2013Flame-propagati,Eiden2020Dynamics-of-Lat}. In contrast, the measured tail luminosity \citep{Casten2023Hydrogen-trigge} is consistent with $Z = 0.02$. Note that $\lesssim\val{500}{\second}$ of the tail were observed \citep{Casten2023Hydrogen-trigge}, however, which is far less than the expected tail duration.

For $Z\geq 0.02$, once the hydrogen in the accumulated layer is depleted, the burning shuts down and the envelope cools until sufficient hydrogen has again accreted to trigger an instability. The burning is then in a limit cycle that evolves on the secular timescale for the underlying neutron star ocean to warm. For $Z = 0.01$, the envelope mass increases during the long tail, which causes $L$ to rise. As a result of this heating, the hot CNO burning does not shut down at the conclusion of the burst, but rather decreases to the lower steady-state burning rate $L_{\mathrm{SS}}$ (Eq.~[\ref{e.Lsteady}]). 

\begin{figure}[htbp]
\centering
\includegraphics[width=\figcolwidth]{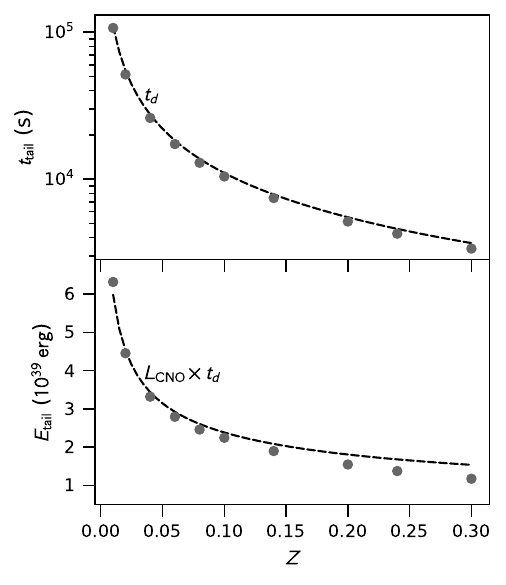}
\caption{\label{fig:tailparams} Duration \ttail\ (\emph{top}) and total radiated energy \Etail\ (\emph{bottom}) of the burst tail as functions of the metallicity $Z$. For comparison, we also show (\emph{dashed lines}) the expected duration of hot CNO burning, $t_{d}$, Eq.~(\ref{e.num-depletion-time}), and $\Lcno\times t_{d}$, where we compute \Lcno\ using Eq.~(\ref{e.LCNO}) and the fit for ignition column (Fig.~\ref{fig:column}). 
}
\end{figure}

The duration, and hence total energy radiated, of the tail depends on the hydrogen-to-CNO ratio (Eq.~[\ref{e.depletion-time}]). To check this scaling, we separate the lightcurves into peak and tail segments using the criteria described in \S~\ref{s.onset-hydrogen-burning}. We then integrate the luminosity to find the total energy radiated in the tail (\Etail). Fig.~\ref{fig:tailparams} displays \ttail\ (\emph{top}) and \Etail\ (\emph{bottom}). The burst models are in good agreement with our estimates (\emph{dashed lines}) for $t_{d}$ (Eq.~[\ref{e.num-depletion-time}]) and $\Lcno\times t_{d}$, where we compute \Lcno\ using Eq.~([\ref{e.LCNO}]) combined with the power-law fit (Fig.~\ref{fig:column}) for \yign. Our estimate for \Etail\ differs from the \MESA\ calculations at both low and high $Z$.  At low $Z$, Eq.~(\ref{e.num-depletion-time}) leads to a slight underestimate (5\%) of \Etail\ because the burst is sufficiently long-lived that the mass of the accreted layer significantly increases over the tail's duration. As $Z$ increases, \ttail\ decreases and the mass accumulated during \ttail\ becomes negligible. The estimate at high $Z$ is 30\% too high, however, because not all of the accumulated hydrogen is consumed. Locating the boundary (defined by where H has decreased to 50\% of its surface abundance) that separates the unburnt hydrogen from the newly burnt ashes indicates that about 90\% of the accumulated hydrogen is consumed during the burst. 

\section{Discussion}\label{s.discussion}

Motivated by the recent discovery \citep{Casten2023Hydrogen-trigge} of a weak flash during a recent outburst of \SAX, we have modeled the envelope of a cool, slowly accreting, neutron star, such as might occur just after the onset of an accretion outburst in a low-mass X-ray binary. Our models of hydrogen ignition span a range of envelope metallicities, and we follow the burst through the thermally unstable ignition of hydrogen and the transition to a steady-state burning via the hot CNO cycle. 
Our main conclusion is that there is indeed a regime of vigorous burning that launches convection and produces a sharp peak, in agreement with previous semi-analytical calculations \citep{Peng2007Sedimentation-a,Cooper&Narayan2007Hydrogen-trigge}. 
As with classical novae, the rapid growth of the instability cuts off after the initial flurry of proton captures locks the CNO nuclei into $\beta^{+}$-decay bottlenecks. Producing a strong burst therefore requires enhanced CNO abundances. For our suite of models, we find that convection, and hence a well-defined, bright peak, only occurs for $Z \gtrsim 0.06$. At lower $Z$, the envelope heats on a thermal timescale and eases into a long plateau of hot CNO cycle burning. Without vigorous convection the flame front would not rapidly spread over the neutron star surface \citep{Cavecchi2013Flame-propagati} but perhaps would instead slowly propagate via a ring of fire \citep{Bildsten1993Rings-of-Fire:-}. If so, the rise would be on the propagation time of the front and much longer than the rise in our one-dimensional simulations.

For higher-$Z$ bursts, the initial, rapid peak in emission only consumes a small amount of H. Following the peak the burning settles into a quasi-steady-state with a luminosity set by the ignition column \yign\ and metallicity, Eq.~(\ref{e.LCNO}). This burning continues until the hydrogen is consumed, the timescale for which depends on the ratio $X/Z$ in the accreted fuel (Eq.~[\ref{e.num-depletion-time}]). At the lowest metallicity ($Z=0.01$) we explored, the burning did not extinguish at the end of the burst, but instead settled into burning hydrogen stably at the rate it accreted, Eq.~(\ref{e.Lsteady}). 

Weak bursts are expected to occur primarily during the early stages of the accretion outburst, when the accretion rate is lower and the surface temperature cooler. In this study, we therefore focused on the first burst following the onset of accretion of hydrogen-rich material onto the neutron star envelope.
The first burst is more energetic than subsequent bursts due to the initially cooler temperature and the longer accumulation time required to reach ignition. As the envelope heats, the bursts are expected to evolve, and we see this in our simulations. For example, with $Z=0.06$ \yign\ decreases by 43\% from the first to the second burst. The morphology of the lightcurve also evolves: during the first burst, the tail luminosity increases (see Fig.\ \ref{f.stacked-bursts}) as the nuclear heating warms the neutron star envelope.  In subsequent bursts, the warmer envelope ensures a relatively constant tail luminosity.

We adopt a specific variation in composition: namely, holding $X$ fixed while varying $Z$ in order to isolate changes in metallicity. In reality, both hydrogen and helium abundances would vary. Models of helium bursts from \SAX\ at higher \Mdot\ are consistent with solar metallicity. By tracking \Mdot\ over an outburst and matching (He) burst recurrence times and fluences, \citet{Johnston2018Simulating-X-ra} estimated $X = 0.44$ and $\Zcno=0.02$, for $d=\val{3.5}{\kpc}$ and $L_{b}/\Mdot = \val{0.30}{\MeV\,\amu^{-1}}$. By fitting a semi-analytical ignition model \citep{cumming.bildsten:rotational} to burst observations, \citet{Goodwin2019A-Bayesian-appr} inferred that $X=0.57$ and $\Zcno=0.013$. These are consistent with the metallicity inferred from the tail luminosity \citep{Casten2023Hydrogen-trigge}, which aligns with our models having $Z=0.02$. In this case, however, our models do not produce a strong peak; matching this peak with our models requires $Z \gtrsim 0.20$. If the weak burst observed by \citet{Casten2023Hydrogen-trigge} is indeed H-powered, then the base of the accreted layer must be greatly enriched relative to that in the accreted material. 

\begin{figure}[htbp]
\centering
\includegraphics[width=\figcolwidth]{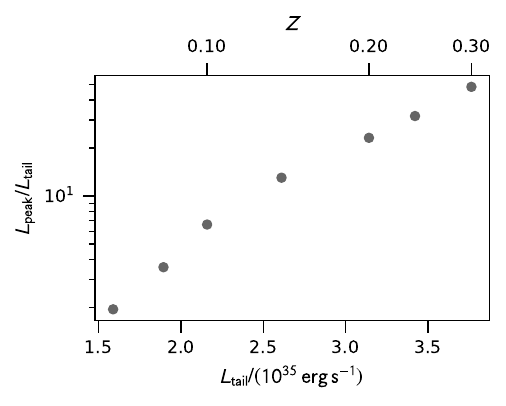}
\caption{\label{fig:obs-ratio} The ratio of \Lpeak\ to \Ltail, measured at the start of the tail. For comparison, we provide the metallicity corresponding to \Ltail\ for our models along the top. 
}
\end{figure}

As described in \S~\ref{s.cno-cycle}, the luminosity and duration of the burst tail depend on the mass and  abundance of hydrogen and CNO catalysts in the accreted fuel. Monitoring of the weak hydrogen flash, including the long tail, can therefore inform us not only about the composition of the accreted material, but also about the extent of sedimentation and mixing in the neutron star envelope. 
Figure~\ref{fig:obs-ratio} displays the distance-independent ratio $\Lpeak/\Ltail$ over a range of \Ltail, which depends on $Z$ (indicated on top axis), both directly, Eq.~(\ref{e.LCNO}), and via the ignition column (Fig.~\ref{fig:column}). Both \Lpeak\ and \Ltail\ are sensitive to the CNO abundance, but at different times: \Lpeak\ is set by the amount of CNO available at the base of the accreted layer during the initial instability; whereas \Ltail\ depends on the available CNO after the envelope has been mixed. Observations of these quantities can therefore inform us not only about the metallicity of the accreted material, but also about the degree of stratification, such as arising from sedimentation or entrainment of previously burned material. Measurements of \Lpeak\ and \Ltail\ from future observations that fall above the points in Fig.~\ref{fig:obs-ratio} would suggest an enhancement of CNO at the base of the layer.

The onset of accretion in an X-ray transient is difficult to predict, and hence the first days of an accretion outburst tend to be sparsely observed. If H-triggered bursts are in fact confined to the few first days of accretion, then it is likely that our sample is quite incomplete. Recently it has become feasible for optical monitoring, by the X-ray Binary New Early Warning System (XB-NEWS), to provide the necessary trigger to catch the X-ray rise of a new outburst \citep{Russell2019Optical-precurs, Goodwin2020Enhanced-optica}. This approach in fact enabled the detection of the 2019 \nicer\ burst and has the potential to aid in observing future additional weak bursts. Obtaining more observations with continuous and complete coverage over the first few days of outbursts from \SAX\ and other X-ray transients could significantly increase the number of detected weak bursts and provide a more detailed picture of the physics in the neutron star envelope.

\begin{acknowledgments}
This work was enabled by the National Science Foundation under grant PHY-1430152 (JINA Center for the Evolution of the Elements), and supported by NASA under grant 80NSSC20K0503. AC and SG were supported by NSERC Discovery Grant RGPIN-2023-03620. It is a pleasure to thank Hendrik Schatz and Laura Chomiuk for numerous discussions on nuclear reactions and classical novae, respectively, Miranda Pikus for assistance on preliminary \MESA\ models, and Duncan Galloway and Tod Strohmayer for a critical reading of the manuscript. AC and EB thank the Institute for Nuclear Theory at the University of Washington for its hospitality and the Department of Energy for partial support during the completion of this work.
\end{acknowledgments}

\software{Modules for Experiments in Stellar Astrophysics 
    \citep[\MESA;][]{Paxton2011, Paxton2013, Paxton2015, Paxton2018, Paxton2019, Jermyn2023},
    \MESA~\code{SDK} \citep{Townsend2024MESA-SDK-for-Li},
    \code{matplotlib} \citep{Hunter2007Matplotlib:-A-2}, \code{numpy} \citep{Harris2020Array-programmi}
}

\bibliography{weakbursts}
\bibliographystyle{aasjournal}

\listofchanges

\end{document}